\documentclass[sigconf,authorversion,natbib=true]{acmart}

\setcopyright{cc}
\setcctype{by}
\copyrightyear{2026}
\acmYear{2026}
\acmConference[arXiv preprint]{}{2026}{}

\acmDOI{}
\acmISBN{}
\renewcommand\footnotetextcopyrightpermission[1]{}

\usepackage{xurl}
\usepackage{booktabs}
\begin{document}

\title[AllSERP]{AllSERP: Exhaustive Per-Element Enrichment of the Versatile AdSERP Dataset}

\author{Andy Edmonds}
\email{andyed@gmail.com}
\affiliation{%
  \institution{}
  \city{}
  \country{}
}

\renewcommand{\shortauthors}{Edmonds}

\begin{abstract}
AdSERP~\cite{latifzadeh2025adserp} is a public corpus of 2,776 search trials with eye gaze, cursor, scroll, pupil and click telemetry recorded on real Google result pages, together with the page screenshots and captured HTML. Its shipped bounding boxes cover only advertisements, which receive about 16\,\% of main-axis clicks. AllSERP adds a typed area-of-interest (AOI) layer for everything else: screenshot-anchored boxes for organic results, People-Also-Ask, image packs, local packs and other widgets, labelled from the captured HTML. Three typed flavours trade box tightness for coverage, from tight boxes to gap-filled boxes to per-card cells inside the top-ads carousel. The ad partition agrees with the shipped ad rectangles on all 38,250 comparisons, and box fidelity is measured on the full corpus with a DOM-based harness. Under the gap-filled flavour, 95.7\,\% of final clicks land in a typed main-axis AOI once cursor coordinates are converted into screenshot space. In a per-element inventory of clicks, fixations, regressions and above-fold exposure, attention and clicks dissociate by element type: 99.7\,\% of top-ads carousels are fixated but only 10\,\% of clicks land on them, and 44\,\% of People-Also-Ask boxes are fixated but only two clicks in the corpus land on one. We release the pipeline, per-trial JSON, corpus CSVs and a browser-based replay viewer.
\end{abstract}

\ccsdesc[500]{Information systems~Users and interactive retrieval}
\ccsdesc[300]{Information systems~Web log analysis}
\ccsdesc[300]{Human-centered computing~User studies}

\keywords{search engine results pages, multimodal datasets, eye tracking, mouse cursor tracking, area of interest, dataset enrichment}

\maketitle

\section{Introduction}\label{sec:intro}

AdSERP~\cite{latifzadeh2025adserp} records what people looked at, moved toward and clicked on real commercial-intent Google result pages. Each of its 2,776 trials carries a full-page screenshot, the captured HTML, 150~Hz Gazepoint eye tracking, \texttt{evtrack} cursor telemetry~\cite{leiva2013evtrack}, scroll and pupil signals. Early work on it predicts per-slot attention from cursor trajectories~\cite{villaizan2025adsight} and develops real-time pupillometric load measures~\cite{duchowski2026rtlfhf,jayawardena2025ripa2}.

The shipped geometry stops at advertisements. AdSERP's bounding boxes cover the top-ads carousel, in-column native ads and right-rail ads, and those surfaces take about 16\,\% of the clicks that land on the main column (about 19\,\% of final clicks once right-rail ads are counted). The rest land on organic results, People-Also-Ask, image packs, local packs and other widgets that have no boxes at all. Analyses that want a per-element read have had to pool ads with organics under an absolute rank, or estimate organic positions from heading counts, which assumes uniform result heights when real results run 80 to 320~px tall.

AllSERP fills that gap. It anchors boxes to the shipped screenshots, labels each box from the captured HTML, and releases three typed flavours, an untyped organic/ad comparator, and per-trial JSON. This paper describes the pipeline, what has been checked about it, what the corpus looks like per element type, how far coverage now reaches, and where it stops.

\noindent\textbf{Contributions.}
\begin{enumerate}
  \item A screenshot-anchored, HTML-typed AOI pipeline covering eight main-axis element types and the off-axis surfaces (\S\ref{sec:pipeline}), released in three typed flavours: tight boxes, gap-filled boxes, and cell-aware boxes that subdivide the top-ads carousel.
  \item Validation against the shipped ad rectangles (38,250 comparisons, no disagreements) and a DOM-based fidelity harness that scores the released boxes, trial by trial, on the full corpus (\S\ref{sec:validation}).
  \item A per-element inventory of click share, fixation coverage, regression rate and above-fold exposure, re-derived on the current substrate (\S\ref{sec:inventory}).
  \item An account of how coverage has grown, across element types and within them, with the maturity of each layer stated so consumers can pick what to trust (\S\ref{sec:coverage}).
  \item Code under MIT, derived data under CC-BY-4.0, and a replay viewer that renders the enrichment on the source screenshots (\S\ref{sec:release}).
\end{enumerate}

\section{The Pipeline}\label{sec:pipeline}

\begin{figure*}[t]
  \centering
  \includegraphics[width=\textwidth]{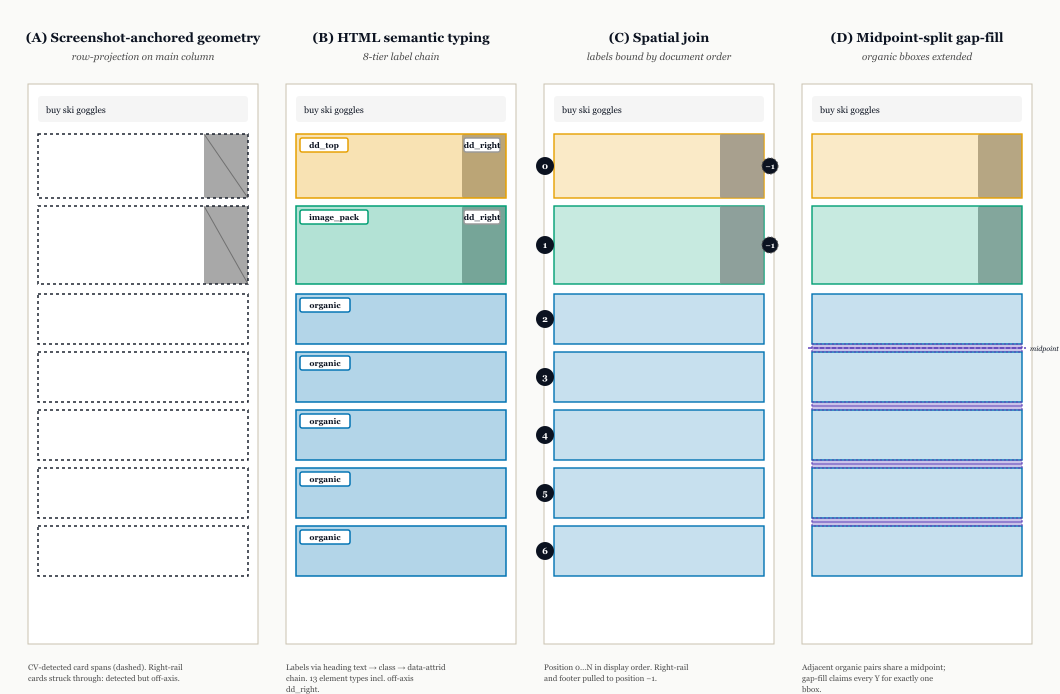}
  \Description[Four-panel pipeline diagram on a synthetic SERP]{Four panels show the same synthetic Google SERP with progressive overlays: Phase A draws dashed brackets around CV-detected card spans; Phase B colours each card with its element type and adds a labelled chip; Phase C numbers each main-axis card 0 through 6 in display order with right-rail cards marked as off-axis; Phase D adds purple shaded gap-fill regions between adjacent organic results, with a midpoint annotation marking the boundary.}
  \caption{The four phases on a synthetic SERP. \textbf{(A)} Row projection on the main column yields card spans. Right-rail cards are detected but routed off-axis. \textbf{(B)} Each card receives a label from the captured HTML. \textbf{(C)} Labels bind to geometry in document order. Main-axis cards get positions $0 \dots N$; right-rail and footer cards get $-1$. \textbf{(D)} Adjacent organic boxes extend to their shared midpoint so every $y$ in the column belongs to one box. Ads and widgets are unchanged.}
  \label{fig:pipeline}
\end{figure*}

\subsection{Why the screenshot, not the DOM}

The AdSERP team extracted their ad boxes from the live DOM at collection time, which is the cleanest source while the page is live~\cite{latifzadeh2025adserp}. The live-DOM route fails on reuse. Re-rendering the saved 2022--2023 HTML in a 2026 browser drifts by about 13~px at the median and about 45~px at the page bottom relative to the original screenshots, which were captured on Chrome~110 on Windows. Missing assets, browser versions, fonts and Google's own layout experiments all contribute. The fixation and cursor streams are coordinated against the screenshots, so AllSERP measures geometry on the screenshots and uses the HTML only for labels.

\subsection{Four phases}

\textbf{Phase A} finds card spans by per-row standard-deviation projection on the main column. Shipped ad rectangles take precedence where they overlap, and a composite-height trigger subdivides image packs, People-Also-Ask and top-stories blocks. \textbf{Phase B} walks the saved HTML and assigns each card a type through an eight-tier priority chain, from heading text through structural class signatures and \path{data-attrid} values to fallback patterns. \textbf{Phase C} binds labels to geometry by document order, the one axis the HTML and the screenshot share, then verifies the pairing geometrically (\S\ref{sec:coverage} describes what that verification changed). \textbf{Phase D} extends each pair of adjacent organic boxes to their shared midpoint, clamped so an organic never crosses an ad or a widget. The midpoint is a heuristic for the boundary between two results, not DOM truth. Misattribution is bounded by the inter-result gap, typically 5 to 60~px, and both the tight and the gap-filled boxes are released so a consumer can pick.

The main-axis label set is \texttt{organic}, \path{dd_top} (top-ads carousel), \path{native_ad}, \texttt{paa}, \path{image_pack}, \path{top_places} (local pack), \path{other_widget} and \path{unknown_widget}. Right-rail ads (\path{dd_right}), right-rail knowledge panels, related searches and page chrome are labelled but carry position $-1$. Widget blocks are first-class slots in display order: a local pack occupies a main-axis position, and 353 of the 2,776 trials (12.7\,\%) contain one. The \path{organic_rank} column numbers organics only, so a widget slot never shifts an organic's rank. On the shipped maps, zero organics are mis-shifted. Each of those 353 trials holds exactly one local pack. Twelve of them are on the exclusion list and one has no fixations, which leaves the 340 local-pack AOIs of Table~\ref{tab:inventory}.

\subsection{Released flavours}\label{sec:flavours}

\begin{table}[t]
  \caption{Released flavours of the corpus CSV. Each is one file with the same schema.}
  \label{tab:flavours}
  \small
  \begin{tabular}{p{2.3cm}p{5.6cm}}
    \toprule
    Flavour & What it is, and when to use it \\
    \midrule
    \texttt{typed} & Phases A to C. Tight boxes on visible text. Use when a fixation must be attributed conservatively. \\
    \addlinespace
    \path{typed_gapfill} & Phases A to D. Organic boxes extended to their midpoints. \textbf{Recommended default} and the basis of \S\ref{sec:inventory}. \\
    \addlinespace
    \path{typed_gapfill_cellsplit} & Superset of the gap-fill flavour. Adds \texttt{role = cell} rows that subdivide the top-ads carousel into its cards, sparse organic sub-cells, and the off-axis right-rail block as a covariate. Filtering \texttt{role = parent} and \path{main_axis} recovers \path{typed_gapfill} exactly. \\
    \addlinespace
    \path{organic_hybrid} & Three labels only (\texttt{organic}, \path{native_ad}, \path{dd_top}) with all-main-axis position numbering. Built from Phase~A geometry and the shipped ad rectangles, so it does not depend on the HTML typing layer. Use for continuous position-rank analyses where the topmost SERP slot is the unit. \\
    \bottomrule
  \end{tabular}
\end{table}

Table~\ref{tab:flavours} lists the three typed flavours and the \path{organic_hybrid} comparator. The cell-aware flavour applies on the $x$ axis what Phase~D applies on $y$: a horizontal carousel's cells are widened across their inter-card gaps so a click in a card margin attributes to the nearest cell rather than being lost.

\subsection{Click attribution and coordinate spaces}\label{sec:clicks}

A click is attributed to a main-axis AOI when both its $x$ and $y$ fall inside the box, with a $\pm 5$~px $x$ and $\pm 10$~px $y$ tolerance for link padding. Requiring $x$ matters: $y$-only attribution routes right-rail ads, page chrome and far-off-target clicks into whichever main-axis box shares their $y$. A trial-level helper, \path{is_main_axis_click}, reports whether a trial's final click lands in a main-axis AOI under this rule, and producers that compute click-outcome features drop the trials it flags.

The two instruments do not write into the same space. AdSERP records three widths per trial (screen 1280, document 1403, window 1422~px). Gazepoint is hardware and reports screenshot space, so fixations need no conversion. \texttt{evtrack} is JavaScript and reports document space, so cursor and click samples do. The released filter, called with its default, compares unconverted \texttt{evtrack} coordinates against screenshot-space boxes. On that path, 237 trials are flagged: 67 final clicks inside a shipped right-rail ad rectangle, 169 on page chrome, search tools or far off any box, and 1 with no usable click, for a corpus-level attribution rate of 91.5\,\% (2,539 of 2,776). Converting each final click through its own trial's document-to-screenshot ratios before the containment test recovers 118 of the flagged trials and loses none, for \textbf{95.7\,\%} (2,657 of 2,776). The chrome and off-target category falls from 169 to 15, while right-rail ad clicks rise from 67 to 103: 36 of the supposed off-target clicks were right-rail clicks whose unscaled $x$ sat inside the main column's apparent range. No trial that attributes in document space fails in screenshot space, so the conversion only corrects.

The conversion also changes the case for gap-fill. Under tight boxes and document-space clicks, 23.9\,\% of approached-and-clicked records had their click land outside the box that was approached. In screenshot space that contamination is 3.9\,\% (94 of 2,406 records). The 510 final clicks once classed as near misses at a box edge drop to 5. At most 36 of them moved to the right rail, so at least 469 were clicks inside a main-axis box that the unscaled $y$ had placed between boxes. The midpoint split still recovers clicks that land in the padding between results, but most of the problem it was built to fix was a coordinate bug.

All click numbers in this paper are reported in screenshot space unless a sentence says otherwise. The \path{is_main_axis_click} helper takes the coordinate space as an argument. Its default stays document space, so that released click-outcome populations reproduce exactly. Called with screenshot space, it admits the 2,657 trials above. The conversion adds trials and removes none, so a producer that switches gains the 118 recovered trials and keeps every trial it had.

\section{Validation}\label{sec:validation}

\subsection{Agreement with the shipped ad rectangles}

The shipped ad rectangles were extracted by the AdSERP authors against the same screenshots the gaze and cursor streams were recorded against, so they are a usable gold standard for the ad-versus-non-ad partition. Two checks run against them. None of the 26,590 Phase~A organic boxes meets the overlap criterion against a shipped ad rectangle: IoU $\geq 0.30$ or intersection covering at least half of either box. Every one of the 11,660 shipped ad rectangles (9,217 native ads, 1,582 top-ads carousels, 861 right-rail ads) appears as a typed entry of the matching label, and every typed ad entry matches a shipped rectangle, with mean IoU 1.000. Together these are 38,250 classifications with no disagreements. The check is bounding box against bounding box and admits no cursor coordinate, so it is indifferent to the coordinate question of \S\ref{sec:clicks}, and it holds unchanged on the current substrate. It is an internal-consistency check, not independent annotation: Phase~C inherits ad identity by spatial overlap and is then compared to the same rectangles. The non-ad partition (organic versus widget types) has no external label set. It is checked against the HTML structure the typing chain consumes, against the DOM harness below, and by visual inspection in the replay viewer. Independent annotation of a held-out subset remains future work.

\subsection{DOM fidelity harness}

A harness (\path{aoi_fidelity.py}) renders each saved SERP, resolves each released box to the DOM element it claims, and scores three per-trial checks over the full corpus. The render is subject to the drift of \S\ref{sec:pipeline}, so these are agreement measures against re-rendered HTML, not independent annotation-accuracy estimates:
\begin{itemize}
  \item the recorded final click falls inside the element its xpath names on 87.7\,\% of trials (2,433 of 2,775);
  \item the median IoU across resolvable box--DOM pairs within a trial is at least 0.5 on 90.8\,\% of trials (2,520 of 2,776); the median of these trial medians is 0.879;
  \item the visible top-carousel card count agrees with the cell export on 29.5\,\% of carousels (466 of 1,582), the weakest layer, discussed in \S\ref{sec:coverage}.
\end{itemize}
All three rates are full-corpus. Before the card-collision fix of \S\ref{sec:coverage} changed 454 maps, the box rate was 93.4\,\%. No full-corpus pre-fix output survives to diff against, so the gap is disclosed, not explained.

\subsection{Gaze--cursor registration}

For each trial with a final click and at least one fixation midpoint in the 1,500~ms before it ($n = 2{,}752$), we measure the minimum distance from those fixations to the screenshot-space final click. The median is 75.7~px (interquartile range 39.1 to 164.0~px, 95th percentile 426.5~px); 407 trials (14.8\,\%) have no fixation within 250~px in that window. The 250~px threshold is descriptive: viewing distance is not reported, and this is not a calibration-accuracy test. Gaze and cursor need not move simultaneously on a SERP~\cite{huang2012user}, so we also report the complementary at-click measure: the fixation containing the click time, or the nearest fixation midpoint within $\pm 1$~s when none contains it. Its median is 468.0~px over 2,730 trials, of which 1,672 use the fallback. These diagnostics summarize temporal coupling; neither identifies the cause of a large distance. The earlier document-space calculation (128.8~px lead-window median) is reproduced exactly by the audit script before conversion.
\begin{figure}[t]
  \centering
  \includegraphics[width=\columnwidth,height=0.42\textheight,keepaspectratio]{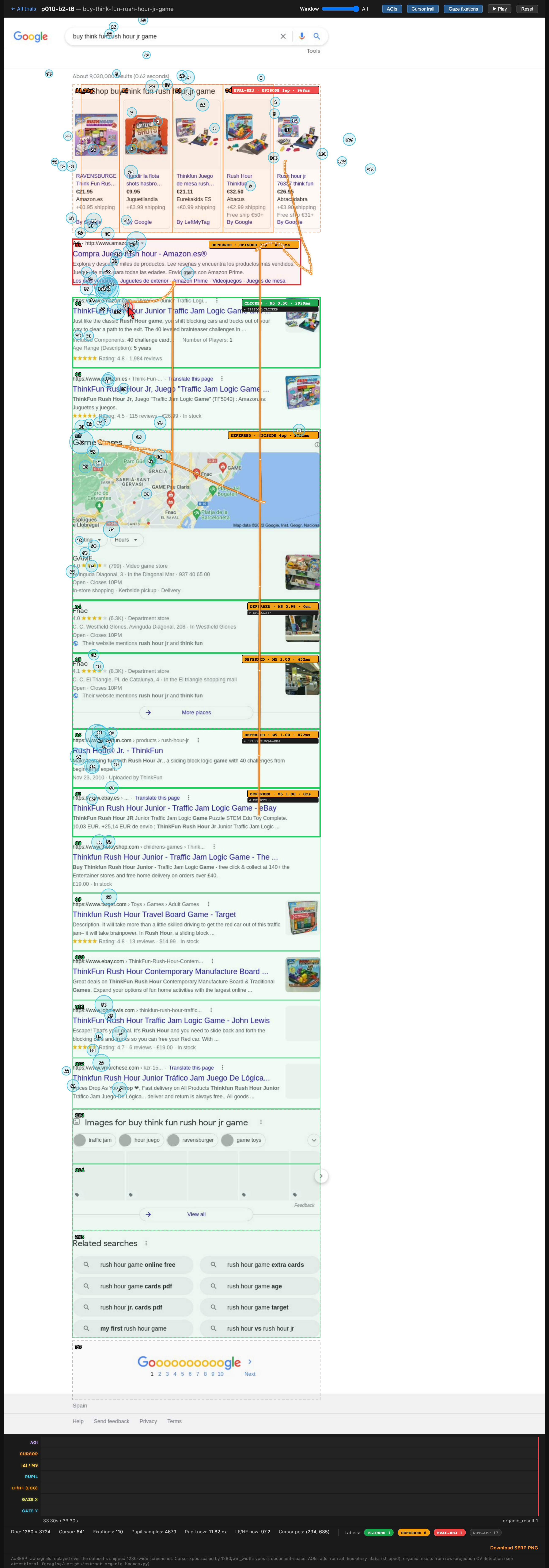}
  \Description[Replay viewer screenshot of one AdSERP trial]{Portrait screenshot of the AllSERP replay viewer: a Google SERP with coloured AOI rectangles, gaze fixation circles, an orange cursor trajectory, and seven labelled timeline tracks at the bottom (AOI presence, cursor, cursor speed, pupil diameter, log LF/HF, gaze X, gaze Y) with min and max value labels.}
  \caption{Replay viewer rendering of trial p010-b2-t6 (\S\ref{sec:release}). Coloured rectangles: typed AOIs with their consideration-set label. Numbered circles: gaze fixations sized by dwell. Orange polyline: cursor trajectory. Timeline tracks: AOI under cursor, cursor position, cursor speed, pupil diameter, pupil LF/HF ratio on a log scale, gaze $x$ and $y$. Each track carries its range in units.}
  \label{fig:replay}
\end{figure}

\section{Element-Type Inventory}\label{sec:inventory}

Table~\ref{tab:inventory} reports the inventory under the gap-fill flavour on the current substrate: 2,764 trials after alignment exclusions (\S\ref{sec:release}), of which 2,763 processed, yielding 36,396 typed main-axis AOIs. Four denominators appear in this paper. Counts against the shipped ad rectangles and the trial-level click filter use all 2,776 AdSERP trials, the fidelity harness uses 2,775 or 2,776 depending on the check, and the typed maps use the 2,764 retained trials, 2,763 once one trial without fixations is dropped. Knowledge panels do not appear because on these pages they render in the right rail and are now routed off-axis (\S\ref{sec:coverage}). Column denominators differ and are given in the notes.

\begin{table*}[t]
  \caption{Per-element-type behavioural inventory under the gap-fill flavour. Element types ordered by AOI count. Click columns use screenshot-space coordinates.}
  \label{tab:inventory}
  \small
  \begin{tabular}{lrrrrrr}
    \toprule
    element type & AOIs & fixated \% & clicks & click \% & regressive \% & above-fold \% \\
    \midrule
    organic         & 22,269 & 55.2          & 2,232 & 81.2 & 56.8 & 93.8 \\
    native\_ad      & 9,180  & 36.4          & 156   & 5.7  & 46.8 & 37.7 \\
    dd\_top         & 1,575  & \textbf{99.7} & 280   & 10.2 & 83.6 & 57.0 \\
    image\_pack     & 1,566  & 50.3          & 40    & 1.5  & 61.7 & 19.2 \\
    paa             & 765    & 43.5          & 2     & 0.1  & 50.8 & 11.0 \\
    unknown\_widget & 673    & 18.1          & 14    & 0.5  & 32.0 & 0.5  \\
    top\_places     & 340    & 91.5          & 24    & 0.9  & 76.2 & 11.8 \\
    other\_widget   & 28     & 39.3          & 1     & 0.0  & 27.3 & 0.3  \\
    \bottomrule
  \end{tabular}
\end{table*}

\paragraph{Column notes} \emph{Fixated~\%} is over the element type's AOIs. \emph{Clicks} counts every click event landing in an AOI of that type, and \emph{click~\%} is that count over the 2,749 clicks attributed to any main-axis AOI in screenshot space (2,627 in document space). The conversion is uneven: 26 of the 29 clicks that document space placed on a People-Also-Ask box move to the organic result directly above it, the one-card shift expected from a 10\,\% $y$ overshoot. The \texttt{evtrack} xpaths of those clicks are organic-result headings, so the converted attribution is the right one. The corpus-level rate in \S\ref{sec:clicks} counts trials whose \emph{final} click lands main-axis under the trial filter, a different denominator. \emph{Regressive~\%} is the share of fixated AOIs that were returned to by a regressive fixation, over the fixated subset, and so is not directly comparable across types whose fixation rates differ widely. \emph{Above-fold~\%} is the share of trials in which at least one AOI of that type sits in the initial viewport.

\paragraph{Reading the table} The top-ads carousel is fixated on nearly every trial it appears in but takes about a tenth of clicks. Native ads show the same shape at lower amplitude, 36\,\% fixated and 6\,\% of clicks. The dissociation between attention and clicks that the original AdSERP analysis reported at the ad-versus-organic level survives at element granularity. Part of the carousel's fixation rate is geometry: it is large, present on 57\,\% of trials, and above the fold on every trial in which it appears, so fixation coverage should be read alongside AOI area rather than as attentional pull. Organic results are above the fold on 94\,\% of trials and take 81\,\% of clicks. Local packs, now typed on 340 AOIs rather than the 84 of earlier releases, are fixated at 91\,\% and returned to regressively at 76\,\%, the highest regressive rate after the carousel. People-Also-Ask is the other extreme: fixated on 44\,\% of its AOIs but clicked twice in the whole corpus once clicks are placed in the maps' coordinate space. On this task it is read, not used.

\begin{figure*}[t]
  \centering
  \includegraphics[width=\textwidth]{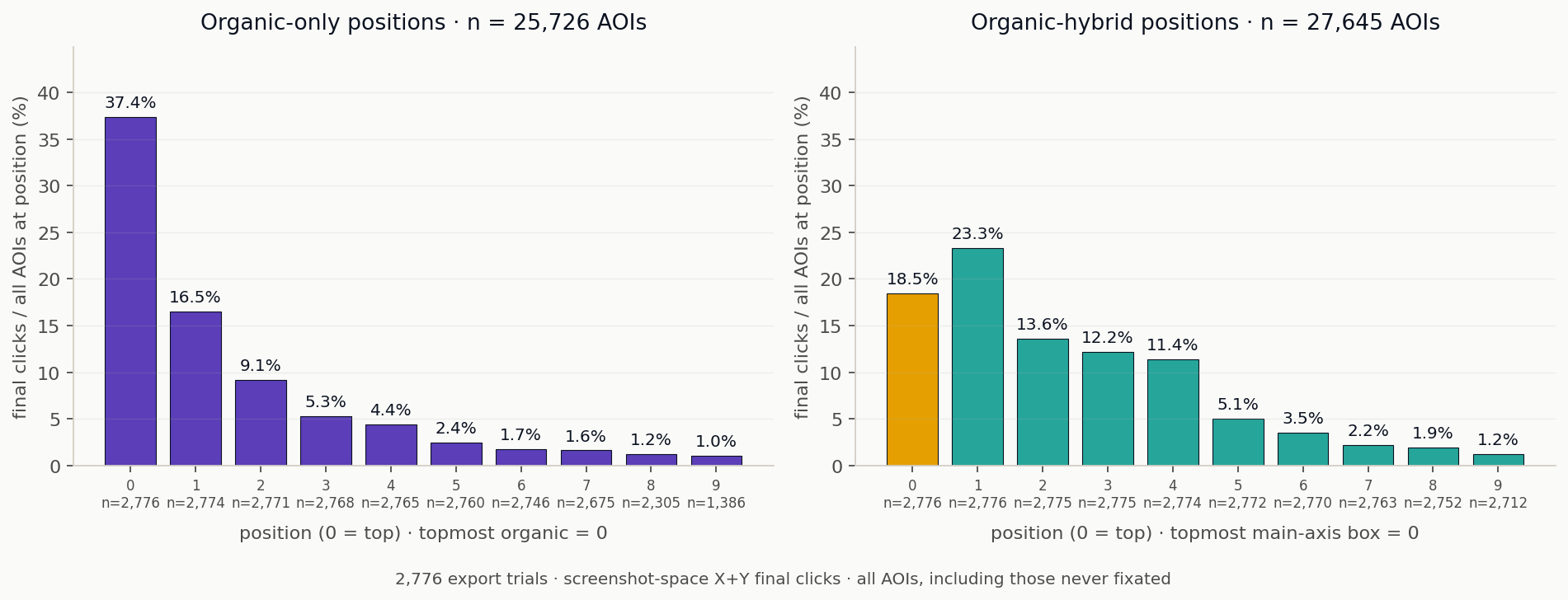}
  \Description[Final-click rates by position with all exported AOIs in the denominator]{Two bar charts show final-click rate by position 0 to 9. Organic-only positions start at 37.4 percent and decline to about 1 percent. All-main-axis positions start at 18.5 percent, peak at 23.3 percent at position 1, then decline. Each bar gives its AOI count; both panels use the same vertical scale.}
  \caption{Final-click rate by position, using all AOIs from the released \texttt{organic\_hybrid} export across 2,776 trials, with no fixation-based selection. Each bar divides attributed final clicks by all exported AOIs at that position; attribution uses screenshot-space $x$ and $y$ without padding. \textbf{Left:} organic rows only, reindexed within each trial (25,726 AOIs at displayed positions 0--9); the first organic has a 37.4\,\% rate. \textbf{Right:} all main-axis rows (27,645 AOIs at displayed positions); position 0 has an 18.5\,\% rate, and position 1 peaks at 23.3\,\%. Across all positions, the organic-only panel attributes 2,224 final clicks and the hybrid panel 2,633. Unmatched clicks and the one trial without a click stay in the AOI denominators. This geometry-only comparison does not use HTML element typing.}
  \label{fig:rank}
\end{figure*}

\begin{figure*}[t]
  \centering
  \includegraphics[width=\textwidth]{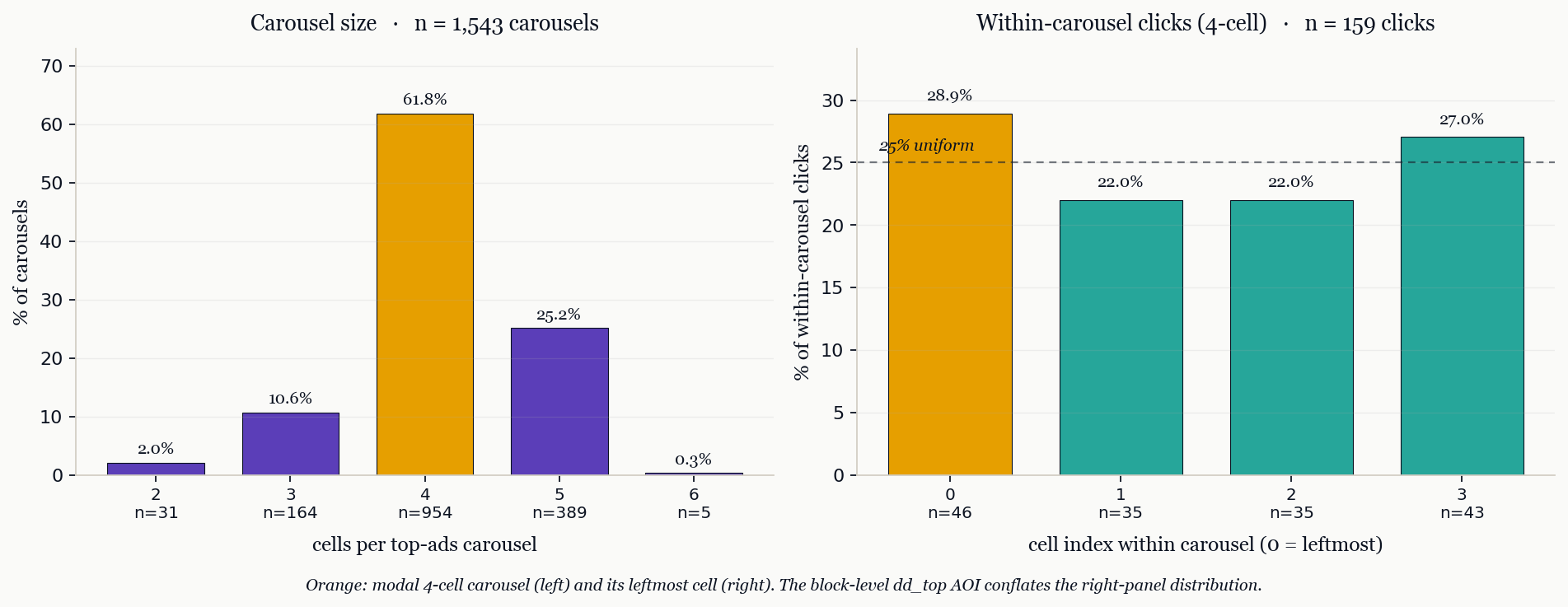}
  \Description[Carousel sizes and final-click composition from the released cellsplit CSV]{Two bar charts. Carousel sizes: 2 cells on 2.0 percent, 3 on 10.6 percent, 4 on 61.8 percent, 5 on 25.2 percent and 6 on 0.3 percent of 1,543 carousels. Final-click shares on four-cell carousels are 28.9, 22.0, 22.0 and 27.0 percent from left to right, over 159 clicks. A dashed line marks 25 percent.}
  \caption{Both panels use the released \texttt{typed\_gapfill\_cellsplit} CSV. \textbf{Left:} cell-count distribution across its 1,543 subdivided carousels. \textbf{Right:} screenshot-space final-click composition on the four-cell layout, 159 of the 264 final clicks inside released carousel cells. Attribution requires strict containment without link-padding tolerance. The 25\,\% line is a descriptive reference, not a significance test.}
  \label{fig:cellsplit}
\end{figure*}

\paragraph{Inside the top-ads carousel} The released cell-aware flavour subdivides 1,543 of the 1,575 retained carousels into 6,345 cells: a median of four cells, range 2 to 6, with four cells on 954 carousels (61.8\,\%). After final clicks are converted to screenshot space, 264 land inside a released carousel cell. On the four-cell layout ($n = 159$ clicks), shares from left to right are 28.9, 22.0, 22.0 and 27.0\,\% (Fig.~\ref{fig:cellsplit}). This is descriptive composition under the released boundaries, not evidence of a monotone position preference. We withdraw the pooled cell-rank click, fixation and dwell endpoints: their producer used a different trial cohort and first-click document-space attribution, so they cannot serve as estimates of the final-click protocol reported here.\footnote{Version~3 reported $\rho = -1.0$ for clicks and $\rho = -0.94$ for fixations and dwell, over six cell means. All three coefficients are withdrawn. The earlier 29.1\,\% cell-agreement figure is also withdrawn because its counter mixed cell types and omitted zero and missing comparisons; the corrected full-corpus figure is 29.5\,\% (\S\ref{sec:validation}).} The cell boundaries still derive from a frozen snapshot with limited card-level agreement (\S\ref{sec:coverage}); pooled rank analyses await the versioned DOM-derived cell source.

\section{Coverage: How Far It Reaches, and Where It Stops}\label{sec:coverage}

Coverage was built in layers, each with a different amount of evidence behind it. Table~\ref{tab:coverage} lists the layers and their status. The caveats come from that history, so this section records it.

\begin{table*}[t]
  \caption{Coverage layers in the release, from the surfaces AdSERP shipped to the ones still being built. Counts are on the current substrate unless a row says snapshot.}
  \label{tab:coverage}
  \small
  \begin{tabular}{p{3.6cm}p{2.2cm}p{3.4cm}p{6.6cm}}
    \toprule
    Layer & Status & Coverage & What certifies it \\
    \midrule
    \multicolumn{4}{l}{\emph{Across element types (one box per card)}} \\
    Ads: carousel, native, right rail & shipped by AdSERP & 11,660 boxes & matched one-to-one by AllSERP, 0 disagreements \\
    Organic results & released & 22,269 main-axis boxes & DOM harness: 90.8\,\% of trials have median IoU $\geq 0.5$ across resolvable pairs, not an organic-only accuracy rate \\
    Widgets: PAA, image packs, local packs, other & released & 3,372 main-axis boxes & HTML structure plus DOM harness. No external label set \\
    Right-rail knowledge panels & off-axis & position $-1$ & screenshot-verified as right rail, not a main-axis slot \\
    \addlinespace
    \multicolumn{4}{l}{\emph{Within element types (cells inside a card)}} \\
    Top-ads carousel cards & released, frozen snapshot & 6,345 cells in 1,543 trials & internal alignment to the parent box only. Visible card count agrees on 29.5\,\% \\
    Top-ads carousel cards, DOM-derived & candidate, not yet released & 7,265 cards, 1,570 of 1,575 subdivisions admitted & screenshot registration of card borders. The 5 rejected cases stay in the denominator \\
    Organic sub-cells & preliminary & 164 cells in 74 trials & 45.8\,\% of candidates align to an organic box. Only the aligned subset ships \\
    Right-rail cells & covariate & 184 cells in 92 trials, 858 blocks & off-axis, shipped so main-axis models can condition on right-rail exposure \\
    Sitelinks and other sub-card structure & not yet & --- & --- \\
    AI Overview cards & not applicable & --- & corpus predates the rollout; a label would be added in Phase~B \\
    \bottomrule
  \end{tabular}
\end{table*}

\paragraph{Across element types} Coverage grew in steps. AdSERP shipped ads. The first AllSERP layer added organic boxes from the screenshots. The second added HTML typing for the widgets that had until then been pooled with organics or dropped. The third added the midpoint gap-fill, after an audit showed that tight boxes lost clicks landing in the padding between results. The fourth, released as AllSERP v1.1.0, replaced index-based label-to-box pairing with geometric verification. Two errors in the earlier maps came to light. Local packs render outside the main results container, so the typing pass never emitted a card for them while the box extractor did, and the box took the next card's label, pushing every card below it one slot down. Right-rail knowledge panels, which render at $x \approx 804$ on a 1,280~px page, had been given a main-column slot they never visually occupied. Correcting both changed the main-axis label sequence on 42\,\% of trials, moved local packs from 84 typed boxes to 340, removed 746 knowledge-panel rows from the main axis, and excluded 12 trials whose rank lattice cannot be verified geometrically. A card-collision fix on 2026-08-30 then removed 454 boxes that duplicated another card's geometry. Aggregate rank statistics barely moved: per-rank medians of the pupil LF/HF ratio~\cite{duchowski2026rtlfhf}, which run from about 14 to 31, shifted by under 0.6 at every rank. Per-trial labels changed, so downstream analyses must identify their substrate and coordinate space.

\paragraph{Within element types} The top-ads carousel is the first surface subdivided below the card. The released cells come from a snapshot frozen in May 2026, aligned to the parent carousel box but not verified card by card. By the DOM harness's count check, visible card counts agree with the snapshot on 29.5\,\% of carousels, with 961 short and 155 over. A replacement source derives cells from the DOM and registers each card border against the original screenshot. On the full corpus it admits 1,570 of 1,575 eligible carousels with matching independent counts, 7,265 cards in all, and the 5 rejected cases stay in the denominator. It is not yet the released source. Adopting it waits on a versioned cell source, so that the per-cell numbers in \S\ref{sec:inventory} are re-derived rather than silently replaced. Organic sub-cells and right-rail cells are shipped at the maturity Table~\ref{tab:coverage} states. The v1.1.1 export applies the current parent boxes and the 12 exclusions, so it keeps 1,543 of the snapshot's 1,550 subdivided carousels. The composition in this version reads the released cellsplit CSV, retains those 1,543 carousels, and converts final clicks to screenshot space. The earlier pooled cell-rank results are withdrawn pending a versioned-source re-analysis.

\paragraph{Exposure is defined by the viewport} A typical carousel is a horizontal scroller several times wider than its viewport (one audited page: 3,796~px of cards clipped to about 652~px, 27 cells in the DOM, about 5 visible). Each \texttt{evtrack} click records the xpath of the element it hit, which identifies the clicked card independently of where the click fell on screen. We attempted to resolve all 2,889 click xpaths against the captured HTML; 67 remained unresolved. The resolved card clicks were joined to the DOM harness, which lists every card in each carousel with its visibility at page load, hidden cards included. Of 281 clicks on top-ads carousel cards in 275 trials (all click events identified by DOM xpath across the corpus, a distinct protocol from the 264 final clicks geometrically attributed to released cells in \S\ref{sec:inventory}), all 281 were on cards visible at load. Paging does occur. One participant pressed the carousel's Next control and then clicked a card that had been on screen from the start. For the cell layer we therefore treat the cards visible at load as the cards a participant was exposed to. Paging that leaves no click, such as horizontal trackpad scrolling, is not observable in this data, so the audit bounds clicks on paged-in cards, not how often participants paged.

\section{Limitations}\label{sec:limits}

\paragraph{Task design} AdSERP elicits one click per query within a one-minute window (extended to two after a prompt) on a single query phrase per trial. The corpus does not sample reformulation, pagination, abandonment, multi-query sessions, or decision arcs longer than that envelope. The statistics in \S\ref{sec:inventory} describe constrained-choice SERP evaluation, not free search.

\paragraph{A pre-AI-Overviews snapshot} Data were collected between 2022-12-16 and 2023-03-13 (from per-trial timestamps), before Google's AI Overviews rollout (limited test May 2023, broad deployment May 2024). The corpus is therefore a clean pre-rollout baseline, and its value as a reference grows as the live SERP diverges from it. Replications on later SERPs need an \path{ai_overview} label, and the AI answer card is likely to draw clicks and fixations differently.

\paragraph{Heuristics and missing labels} The midpoint split and the cell widening are heuristics with bounded error, and both tight and extended boxes ship. The non-ad partition has no independent annotation. The right rail's organic-style results are not recovered, and one trial's final click lands in the right rail with no shipped rectangle to catch it, a property of the original release.

\section{Release and Data Statement}\label{sec:release}

\subsection{Artifacts}

\begin{itemize}
  \item \textbf{Code.} \path{attentional-foraging} (\url{https://github.com/andyed/attentional-foraging}) and \path{approach-retreat} (\url{https://github.com/andyed/approach-retreat}), MIT licence. Single entry point: \path{scripts/build_aois.py} with \texttt{--flavor <name>}.
  \item \textbf{Derived data (CC-BY-4.0).} The four corpus CSVs of Table~\ref{tab:flavours}, matching per-trial JSON (about 19~MB across the corpus), and a per-(trial, position) content-feature file with lexical, query-overlap and semantic-cosine features. Snippet lexical diversity is correlated with rank, so control for rank before claiming a per-position content effect.
  \item \textbf{Replay viewer.} \url{https://andyed.github.io/approach-retreat/replay/} renders typed AOIs, fixations, cursor and the pupil, LF/HF and gaze timelines on the source screenshots for 148 curated trials (Fig.~\ref{fig:replay}). Two of the 148 are on the alignment-exclusion list and are badged as such.
  \item \textbf{Underlying corpus.} AllSERP requires the AdSERP Zenodo volume (\url{https://zenodo.org/records/15236546}), released CC-BY-4.0. AllSERP does not redistribute the corpus. The replay viewer renders screenshots for its 148 curated trials, and the carousel audit evidence carries card crops and captured card text for the trials it inspected, both under the corpus licence with attribution.
\end{itemize}

The corpus CSV is the join target. Existing AdSERP signal files (cursor approach features, pupil power ratios, saccade orientation) merge on \path{trial_id} and gain a per-AOI \texttt{etype} column without re-collecting data. The repository README carries a short Polars example.

\subsection{Substrate identity}

The maps this paper describes are AllSERP enrichment \textbf{v1.1.1}, pinned by content rather than date: 2,764 analysable trials, an alignment-exclusion list of 12 trials with its rule, \texttt{mean\_residual > 30 OR (mean\_h\_mismatch > 60 AND mean\_residual > 8)}, and the typed-map content hash \texttt{2cb789eb8febd234}. The listed trials are set aside, not relabelled. The typed flavours drop them. \path{organic_hybrid} keeps them, because an unverified rank lattice is a typed-label problem and that flavour uses the boxes alone. Membership changed from 14 entries to 12, so re-read the list rather than diff the count. The release notes record how each export summary states the list. A quick tell for a stale export: earlier maps contain about 746 main-column knowledge-panel rows and 84 local packs, and v1.1.1 contains 0 and 340 respectively.

\subsection{Changes since version 3}

\begin{itemize}
  \item The inventory uses v1.1.1. Table~\ref{tab:inventory} loses its knowledge-panel row and gains 256 local-pack AOIs. Click columns, rank plots, carousel composition and gaze--click distances now use screenshot-space clicks. Rank plots use all exported AOIs rather than only fixated records.
  \item The pooled within-carousel rank endpoints and the three v3 rank coefficients are withdrawn (\S\ref{sec:inventory}). The corrected final-click composition is descriptive and retains the released cell-layer limitations.
  \item The coordinate-space finding of \S\ref{sec:clicks} is new, as are the DOM fidelity harness rates and the coverage account of \S\ref{sec:coverage}. The main-axis trial filter now takes its coordinate space as an argument, and the carousel exposure definition rests on an element-level click audit.
\end{itemize}

\subsection{Reproducing the numbers}

Numerical inputs are produced by scripts in the \path{attentional-foraging} repository. Table~\ref{tab:inventory}: \path{allserp_descriptives.py} with \texttt{--flavor typed\_gapfill --space screenshot}. Trial filter and conversion: \path{audit_trial_filter_space.py}. Carousel exposure: \path{audit_carousel_paging.py}. Right-rail and contamination counts: \path{audit_dd_right.py} and \path{audit_cascade_contamination.py}, each with \path{--space}. Ad-rectangle agreement: \path{validate_typed_ads_vs_shipped.py}. DOM harness: \path{aoi_fidelity.py} with \texttt{--all}, whose full-corpus output ships as \path{aoi_fidelity_full_2026-09-07.json}. Gaze-cursor registration: \path{audit_gaze_cursor_coverage.py}. Cell composition: \path{cellsplit_click_composition.py}. Rank plots: \path{allserp_rank_inventory.py}, with the count audit's evidence under \path{docs/evidence/carousel-2026-09-04/}. The corrected registration and composition producers first reproduce their archived document-space aggregates exactly. Their outputs record this gate, the coordinate space, input hashes and exact screenshot/XML dimensions. Figure renderers in the paper repository consume these outputs.

\subsection{Citation and provenance}

AllSERP ships only AOI labels, geometry and the pipeline. The boxes are pixel-anchored to AdSERP screenshots and indexed by AdSERP trial IDs, so any analysis that joins them to gaze, cursor, scroll or pupil streams needs the AdSERP corpus, and both citations apply~\cite{latifzadeh2025adserp}. Applying the pipeline to a different SERP corpus is the one use for which the AllSERP citation alone suffices. AdSERP collection followed the original team's approved IRB. AllSERP processes only de-identified telemetry and screenshots already released under CC-BY-4.0.

\section{Conclusion}\label{sec:conclusion}

AllSERP gives every card on an AdSERP page a box and a label, checks the ad partition against the shipped rectangles, measures the rest against the DOM, and says which layers are certified and which are not. With element type and geometry joined to the multimodal telemetry, per-element questions become tractable on the existing corpus: cursor approach and retreat per AOI rather than per rank, pupillometric load conditioned on element type~\cite{duchowski2026rtlfhf,jayawardena2025ripa2}, and content features paired with behavioural outcomes on the surface where they occurred, with information-foraging measures~\cite{pirolli1999foraging,fu2007snifact} as one natural lens. The inventory here is the descriptive baseline for that work.

\section{AI Use Disclosure}\label{sec:aiuse}

This paper was drafted and revised with Anthropic's Claude (Opus~4.7 for the initial draft, Opus~4.8 for the cell-split revision, Fable~5.1 for the version~4 restructure, Opus~5.5 for earlier version~4 revisions), and OpenAI's Codex for the coordinate audit and publication preparation. Automated passes checked citations against the bibliography and literature notes, flagged ungrounded comparatives and causal language, and cross-checked numbers between sections, tables and captions. All quantitative claims trace to scripts in the released repository. The author takes full responsibility for the claims, citations and methodological decisions in the paper.

\begin{acks}
Thanks to Peter Dixon-Moses for code review and collaboration on downstream use cases, and to Sara Allawati (RMIT University) for the review of the typed AOI maps that surfaced the local-pack and right-rail mislabelling corrected in v1.1.0.
\end{acks}

\bibliographystyle{ACM-Reference-Format}
\bibliography{bib/allserp}

\end{document}